\def\mono32{$^{12}$CO(3$-$2)}
\def\10co{$^{12}$CO(1$-$0)}
\def\21co{$^{12}$CO(2$-$1)}
\documentclass[12pt,preprint]{aastex}
\usepackage{lscape}
\slugcomment{ApJ Letters, {\em accepted 2008 February 27}. DOI:10.1086/587872}
\shorttitle{Warm molecular gas in spiral galaxies?}
\shortauthors{Galaz et al.}
\begin{document}
\title{\mono32 Emission in Spiral Galaxies: Warm Molecular Gas in Action?}  
\author{Gaspar Galaz\altaffilmark{1},
Paulo Cort\'es\altaffilmark{1,2}, Leonardo Bronfman \altaffilmark{2}, and
Monica Rubio\altaffilmark{2}}
\altaffiltext{1}{Departamento de Astronom\'{\i}a y Astrof\'{\i}sica, Pontificia
  Universidad Cat\'olica de Chile, Casilla 306, Santiago 22,
  Santiago, Chile. ggalaz@astro.puc.cl, pcortes@astro.puc.cl}
\altaffiltext{2}{Departamento de Astronom\'{\i}a, Universidad de Chile, Casilla
  36-D, Santiago, Chile. leo@das.uchile.cl, monica@das.uchile.cl}

\begin{abstract}

Using the APEX sub-millimeter telescope we have investigated the
\mono32 emission in five face-on nearby barred spiral galaxies,
where three of them are high surface brightness galaxies (HSBs) lying
at the Freeman limit, and two are low surface brightness galaxies
(LSBs). We have positive 
detections for two of three HSB spirals and non-detections for
the LSBs. For the galaxies with positive detection (NGC0521 and PGC070519),
the emission is confined to  their bulges, 
with velocity dispersions of $\sim 90$ and $\sim 73$ km s$^{-1}$ and
integrated intensities of 1.20 and 0.76 K km s$^{-1}$, respectively. For
the non-detections, the estimated upper limit for the integrated
intensity is $\sim 0.54$ 
K km s$^{-1}$. With these figures we estimate the H$_2$ masses as well
as the atomic-to-molecular mass ratios. Although all the galaxies are
barred, we observe \mono32 emission only for galaxies with prominent
bars. We speculate that bars could dynamically favor the \mono32
emission, as a second parameter after surface brightness. Therefore,
secular evolution could play a major role in boosting collisional
transitions of molecular gas, such as \mono32, especially in LSBs.  
\end{abstract}


\section{Introduction}

Low surface brightness galaxies (LSBs) remain among the most intriguing
galaxies. Defined as galaxies having disk central surface 
brightness fainter than ($B$) 22.0\footnote{Some authors define the
limit as 23.0 mag arcsec$^{-2}$, for example \citet{impey1997}.} mag
arcsec$^{-2}$, they are a product of low
stellar density. LSBs typically have (1) large amounts of atomic gas in
the form of HI \citep{vanderhulst1993} and (2) low star formation rates
(SFRs). In general they have 
sub-solar metallicity, in agreement with the low SFRs \citep{deblok1998} and
consequently the weak production of metals. Also, LSBs have usually
large mass to light ratios, indicating that disk dynamics is dominated
by significant amounts of dark matter halos. This is consistent with their
flat rotation curves, which in many cases extend to
several times the optical radii. 

A key questions about LSBs is what physical
conditions prevent the gas from forming 
stars. Is this due solely to the fact that the gas is not dense
enough to trigger gravitational collapse and form stars? Is the
amount of molecular gas too low to ensure significant stellar
formation episodes? How different is the CO-to-H$_2$ 
conversion factor X = [N(H$_2$)/W(CO)] in LSBs compared to the value
derived for high surface brightness galaxies (HSBs) in preventing their
H$_2$ from being discovered using CO as a tracer? 
Several studies indicate that $X$ is a function of
metallicity \citep{israel1997}, and thus it should not be a surprise
that in LSBs the conversion factor reaches higher values
than those obtained for HSBs. 

The only way to estimate the amount of
molecular gas (H$_2$) in galaxies is to trace the CO emission. Only a
small number of  
detections of \10co and \21co in LSBs have been achieved
\citep{oneil2000b, matthews2001, oneil2003, matthews2005}. 
However, a different tracer such as \mono32 emitted
usually by warm CO, is necessary to better constrain the gas temperature and 
density \citep{dumke2001, muraoka2006}. Here we report on the
detection of \mono32 in two HSB spirals at the Freeman
limit\footnote{Defined as $\mu_0(B) = 21.65 \pm 0.3$ 
mag arcsec$^{-1}$ \citep{freeman1970}.}  from 
\citet{galaz2002, galaz2006}, and the nondetection of the same 
transition in another HSB and two LSBs. 

The weak \10co emission in LSBs galaxies (see references
above) suggests that we search for other $^{12}$CO lines better suited
for warmer environments. The \mono32 transition 
may be more easily detected, since it
is excited in warm gas (E/k=33.2 K, \citet{meier2001}), which
is probably present in LSBs due to the lower metallicity and lack
of dust which normally shield the UV radiation preventing the
excitation of lower CO transitions (10-30 K). The
caveat is whether the UV radiation destroys 
completely the CO or allows higher energy transitions such as
\mono32. Because of its higher characteristic 
temperature and critical density ($n_{cr} \sim 2 \times 10^3$
cm$^{-3}$), the \mono32 transition may be more sensitive 
to warm and/or dense gas involved in stellar formation, a
key insight that could explain why most LSBs have small 
M$_{H_2}$/M$_{HI}$ ratios.

\section{The sample}

We have observed five spirals from the sample of \cite{galaz2002,
 galaz2006}. All of them are face-on spirals, and therefore the
 APEX beam (of  about 18$\prime\prime$) samples their bulges. Two of the five
 galaxies are {\em bonna fide} LSBs, with disk 
 surface brightnesses fainter than 22.0 mag
 arcsec$^{-2}$ (UGC02921 and UGC02081), and three are HSB spirals
 (NGC0521, PGC070519, and NGC7589: see below for details). 
Four of the selected galaxies have measurable near-IR emission,
and therefore they have a significant population of low-mass
 and/or evolved stellar populations. The criteria to choose galaxies
 were the following:
\begin{enumerate}
\item{\em face-on orientation}.- This minimizes the
 extinction for optical observations, allowing us to identify directly the
 sub-mm line width with the pure gas velocity dispersion. It also minimizes the
 inclination bias on the estimated disk surface brightness. 
\item{\em The HI  mass}.- With the aim of accounting for the atomic gas content as an
 additional variable, we have selected galaxies
 with different HI masses. The HI  masses vary between $7 \times 10^8$
 and $51 \times 10^8$ M$_\odot$.  
\end{enumerate}

Some information worth mentioning about the selected 
galaxies follows.
\begin{itemize}  
\item{\em UGC02081}.- Included in the
\citet{impey1996} catalogue of LSBs and studied by \citet{galaz2002,
galaz2006} (LSB100), it is also in the 
HIPASS Catalogue \citep{meyer2004}. It is at $cz_{hel} = 2616$ km
s$^{-1}$ and has a diameter of 23 kpc. It also exhibits a tiny bar in
the central region. 
It has been studied in many optical and near-IR bands, but was not detected 
by {\em IRAS}. It is not a very massive galaxy in terms of its atomic gas mass
\citep{galaz2002}. Its disk central surface brightness ($B$) of 22.4
mag arcsec$^{-2}$ locates this galaxy at the LSB regime. 
It has a small bulge of about 190 pc scale length, with 
colors $B-R = 1.11$ \citep{galaz2006}. The near-IR color of the bulge
is the same as for the total galaxy, meaning that the bulge stellar populations share 
their properties with those of the disk. 
\item{\em NGC7589}.- Classified as an Sa by \citet{impey1996} [LSB473 in
Galaz et al. (2002, 2006)], it has $\mu_0(B) = 21.51$ mag 
arcsec$^{-2}$ and is therefore an HSB galaxy. Located
at $cz_{helio} = 8938$ km s$^{-1}$, it has a diameter of 36.12 kpc
and has traces of a disk bar.  
The difference between its bulge color ($B-R = 2.0$) and the disk color ($B-R =
1.47$) shows that this galaxy has a metal-rich bulge
compared to the bulge of other spirals \citep{galaz2006}. 
Its near-IR bulge color ($J-K_s = 0.81$) suggests 
that the bulge is more metallic than other bulges in spirals 
\citep{galaz2002} and larger in size, with a scale length of 900
pc ($B$ band). Its HI mass of $51.29 \times
10^8 M_\odot$  makes it $\sim 7$ times more massive than UGC02081. In
spite of its absolute magnitude ($M_B = -19.87$ mag),
the galaxy is not detected by {\em IRAS}.
\item{\em PGC070519}.- LSB 463 in \citet{galaz2002, galaz2006}. This
SBc galaxy ($cz = 5244$ km s$^{-1}$) is about 19 kpc diameter. It has
a disk central SB $\mu_0(B) = 21.7$ mag 
arcsec$^{-2}$ and therefore is also an HSB galaxy. This galaxy has 
an absolute magnitude $M_B = -18.63$ and a color   
$B-R = 1.0$ in the bulge and $B-R = 0.84$ in the overall galaxy. This
implies a bulge with larger metallicity compared to that of the disk 
\citep{galaz2002}, for a galaxy with also a large HI mass ($37.15
\times 10^8 M_\odot$) considering its size. It is not detected by
{\em IRAS}. We note that it also presents a noticeable bar which appears to be
clearly ``melted'' with the bulge.
\item{\em NGC0521}.- This is the largest and brightest spiral in our 
sample ($M_B = -20.11$). Classified as an SBsc(r) ($cz = 5018$ km s$^{-1}$),
has a diameter of $\sim 65$ kpc, twice that of the Milky Way. Its central
disk surface brightness is $\mu_0(B) = 21.7$ mag 
arcsec$^{-2}$, so it is an HSB galaxy. It has visible spiral arms
and a prominent bulge with a scale length 
of 720 pc in the $R$ band. The bulge is quite red ($B-R = 1.64$),
with almost no difference in color with the overall galaxy
(LSB059 in Galaz et al. 2006). The near-IR color $J-K_s = 0.74$
suggests a  metallic bulge. It has a large amount of HI ($43.7 \times
10^8 M_\odot$). However, as discussed below, this is not a large
amount considering the remarkable size of the galaxy. It is detected 
by {\em IRAS} in 60 and 100 $\mu$m, with fluxes of about 0.65 and 3.16 Jy,
respectively. It has a noticeable nuclear bar which does not however
extend too far through the disk. 
\item{\em UGC02921}.- An SAB(s)dm galaxy in \citet{impey1996}
($\mu_0(B) = 23.6$ mag arcsec$^{-2}$), located at $cz_{helio} = 3544$ km
s$^{-1}$), it is an LSB galaxy. It has a diameter of 21 kpc
and a large HI mass ($21.88 \times 10^8 M_\odot$). It present a tiny
bar from which two spiral arms are developed.  
\end{itemize}


\section{Observations with APEX}

For APEX observations we use the heterodyne receiver
APEX-2A (345 GHz), tunable in the frequency range  
$279 - 381$ GHz.  The receiver noise
temperature (about 60-70 K) is fairly constant over the entire tuning
range. The telescope beam size 
at 345 GHz is $\sim 18$ arcsec. We observed the five galaxies between
2006 July and 2007 January, using the chopper wheel calibration
technique \citep{kutner1981}, which 
provides main-beam brightness temperature $T_{MB}$, after dividing by
the main beam efficiency $\eta_{MB} = 0.73$. The typical noise system
temperature during the observation was $T_{sys} \sim 150 - 180$ K. 
The total on source integration time was 2 hr on average, with a
velocity resolution per channel of 0.11 km s$^{-1}$.   
We tuned the receiver to the corresponding redshifted \mono32
line. Data reduction was done with the GILDAS-CLASS package
(http://www.iram.fr/IRAMFR/GILDAS); for each galaxy
all spectra were added and a linear baseline subtracted. 
The final spectrum for each galaxy was smoothed to obtain a velocity 
resolution of $\sim 16$ km s$^{-1}$ and an rms noise temperature
$T_{rms} = 5$ mK. Therefore, the typical noise temperature {\em per
channel} is about $5 \times \sqrt{16/0.11} \sim 60$ mK. 

Figure 1 shows final smoothed and summed spectra. Only for galaxies
NGC0521 and PGC070519 we detected 
\mono 32 emission. A Gaussian fit was applied to each
spectrum and the parameters of the fit are summarized in Table
\ref{values}. Detections are defined for signals above 3$\sigma$,
where $\sigma \sim 5$ mK is the rms noise
temperature. Galaxies NGC7589, UGC02081, and 
UGC02921 do not present \mono32 emissions larger than the detection
threshold.

\section{Analysis and discussion}

Driven by the uncertainty in the CO-to-H$_2$ 
conversion factor for spirals in general, we use an approach 
similar to  that of \citet{oneil2000a}  to compute 
the H$_2$ masses for NGC0521 and PGC070519 and the corresponding upper
limits for NGC7589, UGC02081, and UGC02921. Therefore, we assume a 
\10co to \mono32 ratio of 1. This is a reasonable
approach considering,  for example,  the ratio of about 1.2 
obtained, on average, by \citet{meier2001} for a sample of
dwarf galaxies. \citet{dumke2001} obtained similar figures, with a
(3-2)/(1-0) $\sim 1.3$ for the centers of spirals.  
\begin{figure*}
\plotone{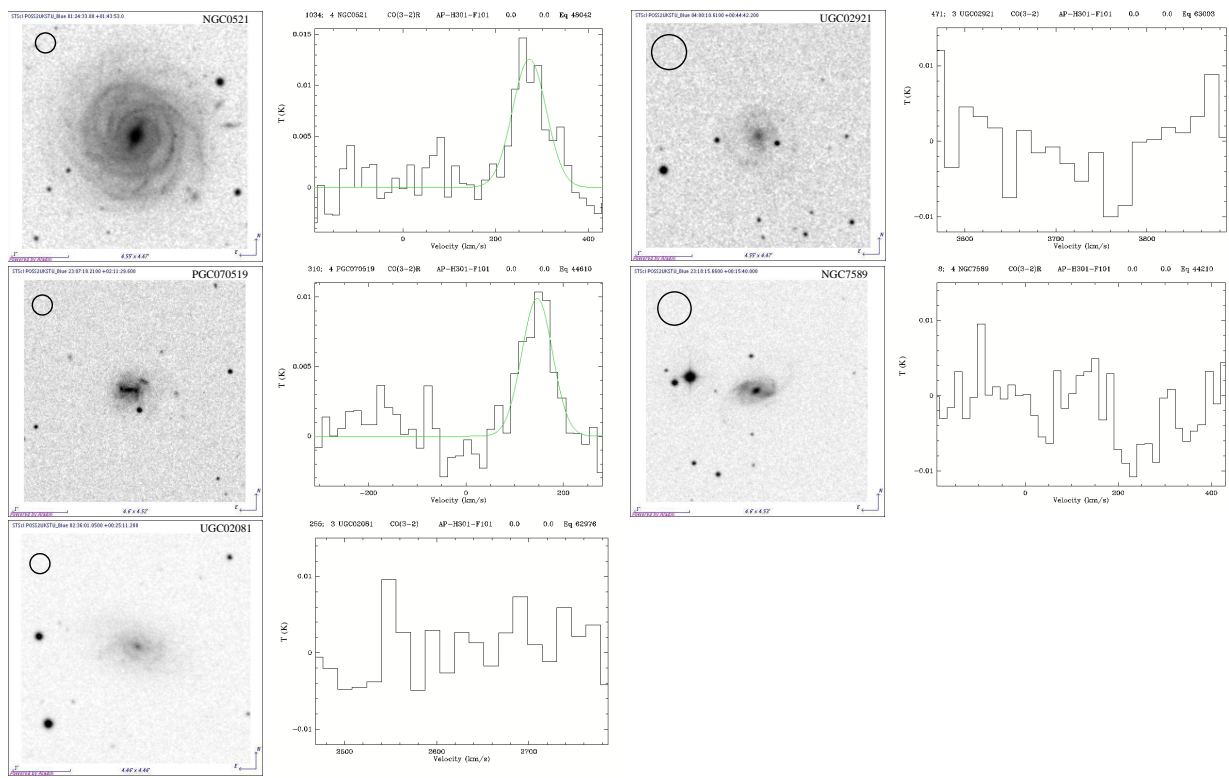}
\figcaption{Images and spectra of galaxies observed with
APEX. Images are are from the SDSS ($B$ filter). Spectra are centered
in the \mono32 rest emission. The velocity 
resolution was smoothed to about 16 km s$^{-1}$ for each spectrum. The solid
line for NGC0521 and PGC070519 spectra indicate the corresponding
Gaussian fits to the line emission. The circle in each image represents
the mean antenna beam size of about 18 arcsec.}
\figurenum{1}
\end{figure*}
CO velocity dispersions are $\sim 89$ km s$^{-1}$ for NGC0521 and 73
km s$^{-1}$ for PGC070519. Since 
these galaxies are face-on, the velocity width would correspond to the 
intrinsic velocity dispersion of the \mono32 emission. In both cases,
and given the similar sizes of the corresponding bulges and
the antenna beam, the velocity dispersions would simply correspond to that of 
the CO embedded in the respective bulges. Note
that the velocity dispersions are $\sim 5 - 10$ times larger than those typically
observed for HI in the central regions of
spirals \citep{vanderkruit1982}. We suspect that this large
width is due to velocity fields induced by the bars observed in
NGC0521 and PGC070519 (see Figure 1). Such behavior was also 
observed by \citet{dumke2001}, who obtained similar values for $\Delta
V$. In Table \ref{values} we indicate the \mono32 line velocity width and
also the HI velocity dispersion. We note the high value also for W(HI)
in NGC0521, strongly indicating the presence of a strong
velocity field. Note that all the other galaxies present large
velocity dispersions in HI, suggesting therefore a likely large value
for the velocity dispersion of their molecular gas. 

From the Gaussian fits we determine the integrated intensity
I$_{CO}=\int{T_{MB}} dv$, and the intensity at the peak of the
emission (in mK, see Table \ref{values}). Using the method of
\citet{bregman1988} and the formula of \citet{sanders1986} 
\begin{equation}
M_T(H_2) = 5.82[\pi/4]d_b^2I_{CO},
\end{equation}
where $d_b$(pc) is the telescope beam diameter at the distance of the
source and $I_{CO}$ is the total CO line integrated intensity (K km
s$^{-1}$), we estimate the H$_2$ mass for galaxies with positive
\mono32 detections (NGC0521 and PGC070519). For galaxies with negative
detections (UGC02081, UGC02921 and NGC7589) we estimate upper limits using  
\begin{equation}
I_{CO} \le 3 T_{MB} \Delta v_{HI}/\sqrt{n},
\end{equation}
\noindent in K km s$^{-1}$. We assume that $\Delta V_{HI} \sim
<\Delta V_{^{12}CO(J=3-2)}>$, the average velocity dispersion between
NGC0521 and PGC070519, that is,
80 km s$^{-1}$. We think this value is more realistic than just using an
arbitrarily smaller value, since UGC02081, UGC02921, and
NGC7589 also present bars. Thus $\sqrt{n}
= \sqrt{80/16}=\sqrt{5}$, the number of 
channels used in  the smoothed spectra. Therefore
\begin{equation}
I_{CO} \le \frac{3}{\sqrt{5}} T_{MB} \Delta v_{CO}.
\end{equation}
For non-detections (UGC02921, UGC02081 and NGC7589),
we use $T_{MB} = \sigma_{rms} \sim 
5$ mK for all the corresponding spectra. Thus, $I_{CO} < 0.54$ K km
s$^{-1}$. Using the corresponding beam  
diameter in pc for all galaxies, as seen at the distance of each
galaxy, we obtain estimated H$_2$ masses for NGC0521 and PGC070519,
and upper limits for UGC02921, UGC02081, and NGC7589 (see eq. [1] and
Table \ref{values}). It is worth noting that we do detect 
\mono32 for NGC0521 and PGC070519, but do not detect it for NGC7589,
which is 0.2 mag 
arcsec$^{-2}$ {\em brighter} than these two galaxies, suggesting
that factors other than surface brightness must play a key role in the
\mono32 emission. 

Table \ref{values} presents $H_2$ masses. Note that
these values are computed assuming that molecular gas is uniformly
distributed through galaxies, which we know is not 
the case, and in all galaxies, the beam was pointed to the bulge. As
shown in Table \ref{values}, H$_2$ masses derived from detections are 
$\sim 10^8$ M$_\odot$. The estimated upper limit for the
molecular mass in the beam size of galaxies with no detections is
$\sim 3-28 \times 10^7$ M$_\odot$. With these values we are
able to compute the molecular-to-atomic gas mass fractions for each 
galaxy (again, only upper limit estimates for galaxies with no
detections). The results agree with the picture that LSBs lack
significant amounts of molecular gas and, in general, with small
molecular-to-atomic mass ratios (below 0.08). For one case (NGC0521)
the 
molecular-to-atomic gas fraction appear as large as 0.9. Although
surprising, this result could be anticipated given its small HI mass
considering its large size. In other terms, the molecular gas  
content for NGC0521 is comparable to its atomic one. Note that we
do not detect \mono32 for three but instead for two HSBs and do not detect any
CO for both of the LSBs. How could one explain this puzzling picture?
Although the central disk SB of these galaxies seems to play a role in
the CO emission, we suspect that other factors are key in
allowing such an emission. Looking in detail at the structure of
the galaxies, we note that NGC0521 and PGC070519 have much more
prominent bars really competing with the bulge emission, compared to
the other three
galaxies (see Fig. 1 and also figures in Galaz et al. 2006), including
NGC7589, which is 0.3 mag arcsec$^{-2}$ {\em brighter} than NGC0521 and
PGC070519. We suggest that strong velocity fields could be
responsible for the molecular emission in LSBs.

\section{Conclusions}

We detect \mono32 in 2 of 5 nearby spirals, using the APEX
submillimeter telescope. The emission is detected for NGC0521 and PGC070519, both
HSB galaxies lying at the Freeman limit. The other galaxies, with no
detections, share similar morphology 
and orientation, but two are LSBs (UGC02921 and UGC02081) and one an
HSB (NGC7589). All of them are
face-on, making the internal extinction negligible. The measured
main-beam temperature CO fluxes  are 1.20 K km s$^{-1}$ for NGC0521
and 
0.76 K km s$^{-1}$ for PGC070519. The remaining galaxies have fluxes
below 3$\sigma$, and thus we are able only to {\em estimate} upper
limits for their \mono32 fluxes ($< 0.54$ K km s$^{-1}$). 

Measured velocity dispersions for NGC0521 and PGC070519 are 89 and 73
km s$^{-1}$, respectively, $\sim 5 - 10$ times typical values
obtained by other authors for bulges in spirals
\citep{vanderkruit1982, dumke2001}. We suspect that in these
cases the gas velocity field is dominated by the bar kinematics
\citep{dumke2001}. Bars are observed actually in all five galaxies, but
those of NGC0521 and PGC070519 appear as the more prominent ones. 

We compute the total H$_2$ mass in the main beam, obtaining $2 \times
10^8 M_\odot$ and  $1.4 \times 10^8 M_\odot$ for NGC0521 and
PGC070519, respectively. The corresponding molecular-to-atomic gas
fraction is about 0.9 and 
0.08. The high value for NGC0521 is probably due to its
low HI density, and to the assumption that the HI is uniformly
distributed over the whole galaxy optical size. 
In fact, in spirals one expects that the HI is mostly
distributed along the disk.  

Overall, we have shown that \mono32
emission could be intense for some galaxies at the SB ``Freeman
limit.'' Moreover, such an emission could be ``dynamically boosted''
by bars, helping to warm the CO and allowing the \mono32
transition at 33 K above the ground level. Although we have no
detections of \mono32 for the LSBs in this sample, we speculate that
many LSBs could present \mono32 emission 
given some dynamical conditions that warm the gas (or make it
denser). We speculate that many LSBs with no \10co or \21co
emission could have instead \mono32 emission thanks to the
poor UV shielding favored by the low metallicity and low dust content,
allowing a higher gas temperature. This emission, which in principle
could be  small, may be largely amplified by secular processes which
warm the gas.   

\acknowledgements 

G.G. acknowledges support from FONDECYT 1040359. P.C. was funded by
the CONICYT-ALMA Fund, project 31050003. L.B. and M.R. acknowledge support from
the Center for Astrophysics FONDAP 15010003. We are grateful to the
anonymous referee for helping to improve this Letter.

\begin{landscape}
\begin{table}
\renewcommand{\arraystretch}{0.6}
\caption{Gaussian fit parameters for \mono32 emission in NGC0521 and
PGC070519, and upper limits for non-detections.} 
\tablewidth{18cm}
\begin{tabular}{lllllllllll}
\hline\hline
Name & $T_{MB}$ & $\sigma_{rms}$ & $\int T_{MB} dv$ & M$_{HI}$ & M$_{H_2}$(18 arcsec) &
     M$_{H_I}$(18 arcsec) & M$_{H_2}$/M$_{HI}$ & Width & W(HI) \\
     & (mK) & (mK) & (K km s$^{-1}$) & ($\times 10^8$M$_\odot$) & ($\times
     10^8$M$_\odot$) & ($\times 10^8$M$_\odot$) & & (km s$^{-1}$) &
     (km s$^{-1}$) \\
(1) & (2) & (3) & (4) & (5) & (6) & (7) & (8) & (9) & (10) \\
\hline
NGC0521   & 12.7  & 2.9 & 1.20 $\pm 0.10$ & 43.7 & 2.01  & 2.2 & 0.91
& 88.7 & 244 \\
PGC070519 &  9.8  & 2.5 & 0.76 $\pm 0.09$ & 37.1 & 1.39  & 16.5 & 0.08
& 72.8 & 64 \\ \hline
NGC7589   & $< 5$ & $\sim 5$ & $< 0.54$        & 51.3 & $< 2.77$  & 26.6 & $<
0.10$ & $\sim 80$ & 183 \\
UGC02081  & $< 5$ & $\sim 5$ & $< 0.54$	    & 7.76 & $< 0.25$ & 1.1 & $<
0.23$ &  $\sim 80$ & 179 \\
UGC02921  & $< 5$ & $\sim 5$ & $< 0.54$	    & 21.8 & $< 0.45$ & 13.4 & $<
0.03$ &  $\sim 80$ & 168 \\ 
\hline
\end{tabular}
\\
{\tiny {\bf Notes}: (1) From \citet{galaz2002, galaz2006} and the NED. (2)
Calibrated peak main-beam brightness temperature. (3) The rms
estimates are from the smoothed data with 16 km s$^{-1}$ resolution. (4) Integrated
intensity. (5) HI mass from \citet{galaz2002} and the NED. (6)
Estimated H$_2$ mass enclosed by the APEX main-beam of 18
arcsec. Values for last 3 galaxies are upper limits. (7) HI mass
interpolated to the APEX main-beam. (8) H$_2$
to HI mass ratio. (9) Velocity width of the line. For the last 3
galaxies the velocity width is estimated as the average value obtained
for NGC0521 and PGC070519. (10) Velocity width for the HI line,
obtained from HIPASS \citep{meyer2004} and HYPERLEDA
\citep{paturel2003}, clipped at 20\% peak flux density. For 
PGC070519, and NGC7589 the indicated value correspond to the mean
homogenized maximum rotation velocity uncorrected for inclination.}
\label{values} 
\end{table}
\end{landscape}
\end{document}